\documentclass[a4paper]{jpconf} 
\usepackage{graphicx} 
\usepackage{bm} 
 
\newcommand{\gapp}{\,{\raisebox{-.2ex}{$\stackrel{>}{_\sim}$}}\,} 
\newcommand{\lapp}{\,{\raisebox{-.2ex}{$\stackrel{<}{_\sim}$}}\,} 
 
\newcommand{\eq}{{\,=\,}} 
\newcommand{\Tc}{T_{\rm cr}} 
\newcommand{\ec}{e_{\rm cr}} 
\renewcommand{\pt}{p_\perp} 
 
\newcommand{\bce}{\begin{center}} 
\newcommand{\ece}{\end{center}} 
\newcommand{\beq}[1]{\begin{eqnarray}\label{#1}}
\newcommand{\eeq}{\end{eqnarray}}

\begin{document} 

\vspace*{-1cm}

\title{Equation of State and Collective Dynamics} 
 
\author{Ulrich Heinz\footnote{$\!\!$Email: {\tt heinz.9@osu.edu}. 
Work supported by the U.S. Department of Energy, grant DE-FG02-01ER41190.}} 
 
\address{Department of Physics, The Ohio State Univesity, 
%191 West Woodruff Avenue, 
Columbus, OH 43210, USA} 
 
%\ead{heinz@mps.ohio-state.edu} 
 
\begin{abstract} 
This talk summarizes the present status of a program to quantitatively 
relate data from the Relativistic Heavy Ion Collider (RHIC) on  
collective expansion flow to the Equation of State (EOS) of hot and  
dense strongly interacting matter, including the quark-gluon plasma  
and the quark-hadron phase transition. The limits reached with the 
present state of the art and the next steps required to make further  
progress will both be discussed.  
\end{abstract} 

\vspace*{-8mm} 
%%%%%%%%%%%%%%%%%%%%%%%%%%%%%%%%%%%%%%%%%%%%%%%%%%%%%%%%%%%%%%%%% 
\section{The QCD Equation of State and ideal fluid dynamics} 
\label{sec1}
%%%%%%%%%%%%%%%%%%%%%%%%%%%%%%%%%%%%%%%%%%%%%%%%%%%%%%%%%%%%%%%%% 
 
\begin{figure}[hb] 
\vspace*{-4mm}
 \bce 
  \includegraphics*[bb=  50 50 410 302,width=0.49\linewidth,height=5cm]% 
  {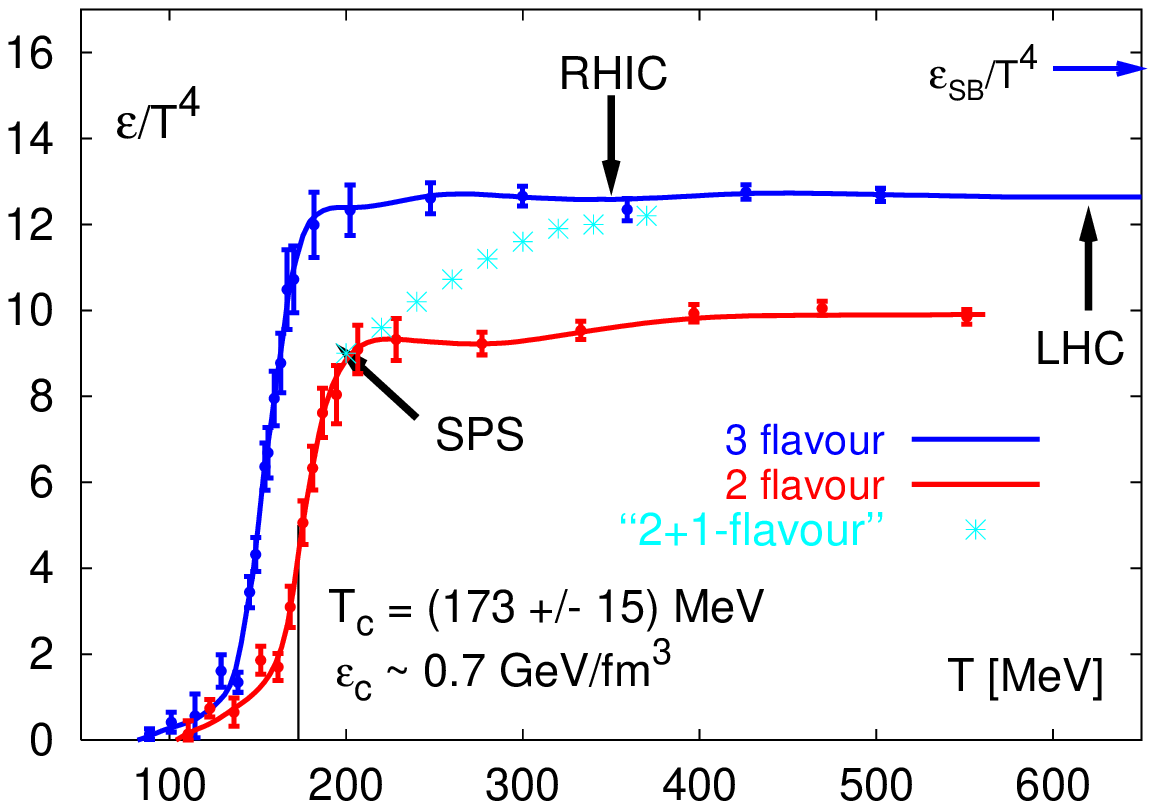} 
  \includegraphics*[bb=  50 50 410 302,width=0.49\linewidth,height=5cm]% 
  {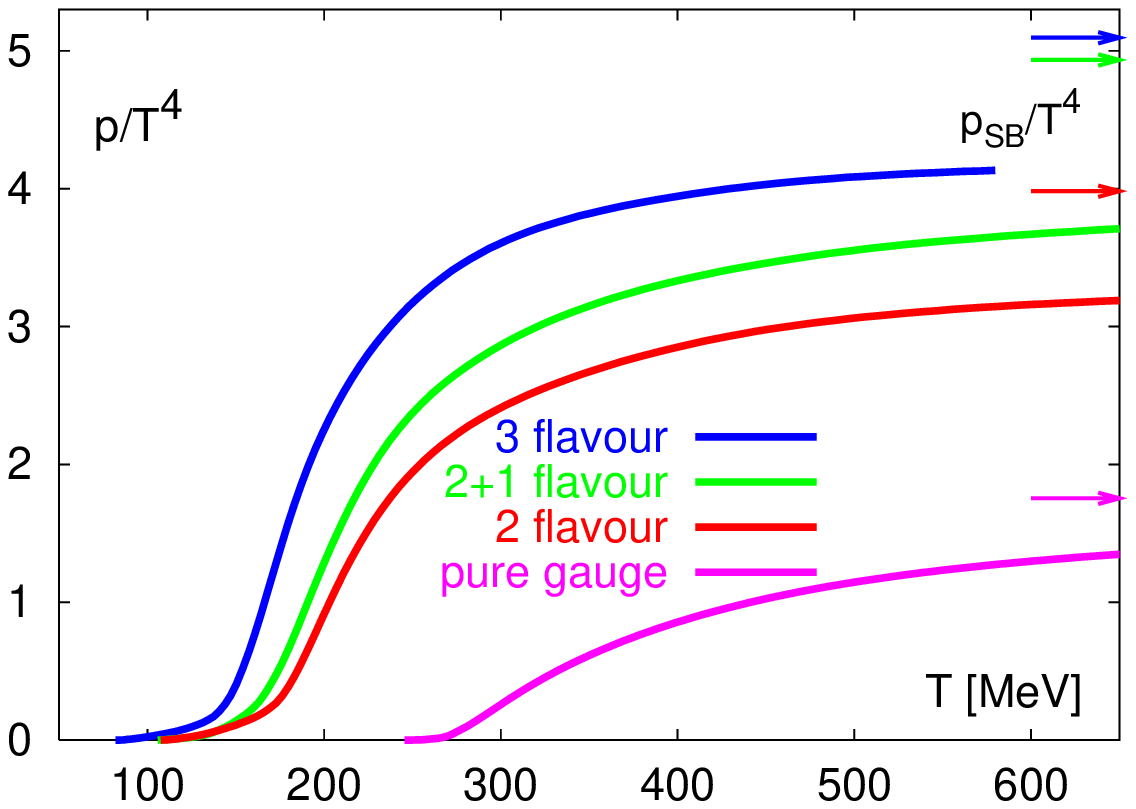} 
 \ece 
\vspace*{-5mm}
 \caption{\label{F1} \small
 The normalized energy density $e/T^4$ (left) 
 and pressure $p/T^4$ (right) from lattice QCD \cite{KL04} for 
 0, 2 and 3 light quark flavors, as well as for 2 light + 
 1 heavier (strange) quark flavors. Horizontal arrows on the 
 right indicate the corresponding Stefan-Boltzmann values for a  
 non-interacting quark-gluon gas.} 
\vspace*{-3mm}
\end{figure}

With relativistic heavy-ion collisions one explores the 
phase diagram of strongly interacting bulk matter in the regime 
of high energy density and temperature. Lattice QCD (LQCD) tells us 
\cite{KL04} that for zero net baryon density QCD matter  
undergoes a phase transition at $\Tc\eq173\pm15$\,MeV from a  
color-confined hadron resonance gas (HG) to a color-deconfined  
quark-gluon plasma (QGP). The critical energy density  
$\ec{\,\simeq\,}0.7$\,GeV/fm$^3$ \cite{KL04} corresponds roughly to  
that in the center of a proton. At the phase transition,  
the normalized energy density $e/T^4$ rises rapidly by about an  
order of magnitude over a narrow temperature interval  
$\Delta T{\,\lapp\,}15-20$\,MeV, whereas the pressure $p/T^4$ (which 
is proportional to the grand canonical thermodynamic potential) 
is continuous and rises more gradually (Fig.~\ref{F1}). Both seem to
saturate at about 80-85\% of the Stefan-Boltzmann value for an ideal
gas of noninteracting quarks and gluons, the energy density more
quickly (at about $1.2\,\Tc$), the pressure more slowly. Above about 
$2\,\Tc$, the lattice data follow the Equation of State of an ideal
gas of massless particles, $e=3p$.
 
For many years this observation has been interpreted as lattice QCD 
support for the hypothesis of a weakly interacting, perturbative QGP. 
The recent RHIC data taught us (as I will show) that this interpretation 
was quite wrong.

It was recognized over 3 decades ago (see review \cite{SG86}) that
%
%%%%%%%%%%%%%%%%%%%%%%%%%%%%%%%% Fig 1 %%%%%%%%%%%%%%%%%%%%%%%%%%%%%%%%% 
%%%%%%%%%%%%%%%%%%%%%%%%%%%%%%%%%%%%%%%%%%%%%%%%%%%%%%%%%%%%%%%%%%%%%%%%%
%
information about the EOS of strongly interacting matter can be extracted
by studying the collective dynamics of relativistic heavy-ion collisions.
This connection is particularly direct in the framework of {\em ideal fluid
dynamics} which becomes applicable if the matter formed in the collision
approaches {\em local thermal equilibrium}. The latter requires sufficiently 
strong interactions in the medium that local relaxation time scales are
shorter than the macroscopic evolution time scale. In this limit the
local conservation laws for the baryon number, energy and momentum currents,
$\partial_\mu j_B^\mu(x)\eq0$ and $\partial_\mu T^{\mu\nu}\eq0$, 
can be rewritten as the relativistc Euler equations for ideal fluid motion:
\beq{e1}
  \dot n_B &=& - n_B\, (\partial\cdot u),\qquad
  \dot e = - (e+p)\, (\partial\cdot u),\\\label{e2}
  \dot u^\mu &=& \frac{\nabla^\mu p}{e+p} 
     = \frac{c_s^2}{1+c_s^2}\,\nabla^\mu\ln\left(\frac{e}{e_0}\right).
\eeq
The dot denotes the time derivative in the local fluid rest
frame ($\dot f\eq{u}\cdot\partial f$) and $\nabla^\mu$ the 
gradient in the directions tranverse to the fluid 4-velocity
$u^\mu$. The first line describes the dilution of baryon and 
energy density due to the local expansion rate $\partial\cdot u$, 
which itself is driven according to (\ref{e2}) by the pressure 
or energy density gradients providing the fluid acceleration. The 
absolute value of the energy density $e$ is locally irrelevant:
the initial maximal energy density $e_0$ only matters by setting 
the overall time scale between the beginning of hydrodynamic 
expansion and final decoupling, thereby controlling how much flow 
can develop globally. The details of the flow pattern are thus 
entirely controlled by the temperature dependent 
speed of sound $c_s^2\eq\frac{\partial p}{\partial e}$.  

According to the LQCD data, the latter is $c_s^2{\,\approx\,}\frac{1}{3}$ 
for $T{\,>\,}2\,\Tc$, then drops steeply near $T{\,\approx\,}\Tc$ to 
values near $c_s^2{\,\approx\,}\frac{1}{20}$ (the ``softest point'', see 
Fig.~11 in \cite{KL04}), before rising again in the hadron resonance gas 
phase to $c_s^2{\,\approx\,}0.15$ \cite{SHKRPV97}. A key goal of flow 
studies in relativistic heavy ion collisions is to find traces of this 
``softest point'' in the data.

\vspace*{-3mm}
%%%%%%%%%%%%%%%%%%%%%%%%%%%%%%%%%%%%%%%%%%%%%%%%%%%%%%%%%%%%%%%%% 
\section{``Flavors'' of transverse flow in heavy ion collisions} 
\label{sec2}
%%%%%%%%%%%%%%%%%%%%%%%%%%%%%%%%%%%%%%%%%%%%%%%%%%%%%%%%%%%%%%%%% 
 
Experimentally one studies flow by analyzing the transverse
momentum spectra of the emitted hadrons. In central ($b\eq0$) collisions 
between spherical nuclei, the flow is azimuthally symmetric about
the beam axis. This ``radial flow'' integrates over the entire pressure 
history of the collision from initial thermalization to final decoupling 
(``freeze-out''), due to persistent pressure gradients.
In noncentral ($b{\,\ne\,}0$) collisions, or central collisions between
deformed nuclei such as uranium \cite{HK05}, the nuclear reaction zone
is spatially deformed, and anisotropies of the transverse pressure 
gradients result in transverse flow anisotropies. These can be quantified
by Fourier expanding the measured final momentum spectrum 
$dN/(dy\,\pt d\pt\,d\phi_p)$ with respect to the azimuthal angle $\phi_p$.
For collisions between equal nuclei, the first non-vanishing Fourier
coefficient at midrapidity is the {\em elliptic flow} $v_2(\pt,b)$. Since 
$v_2$ is driven by pressure {\em anisotropies} and the spatial deformation 
of the reaction zone creating such anisotropies quickly decreases as
time proceeds, the elliptic flow is sensitive to the EOS only during 
the early expansion stage \cite{Sorge97} (the first $\sim5$\,fm/$c$ in
semicentral Au+Au collisions \cite{KSH00}), until the spatial deformation
has disappeared. 

Depending on the initial energy density (i.e. beam energy), the hot 
expanding fireball spends this crucial time either entirely in the QGP 
phase, or mostly near the quark-hadron phase transition, or 
predominantly in the hadron resonance gas phase \cite{KSH00}, thereby
probing different effective values of the sound speed $c_s$. To the 
extent that ideal fluid dynamics is valid in all these cases, an 
excitation function of the elliptic flow $v_2$ should thus allow to
map the temperature dependence of the speed of sound and identify
the quark-hadron phase transition, via a minimum in the function
$v_2(\sqrt{s})$ \cite{KSH00}. This will be further discussed below
(see Section \ref{sec5b} and Fig.~\ref{F6}).

\vspace*{-3mm}
%%%%%%%%%%%%%%%%%%%%%%%%%%%%%%%%%%%%%%%%%%%%%%%%%%%%%%%%%%%%%%%%% 
\section{Model parameters and predictive power of hydrodynamics}
\label{sec3}
%%%%%%%%%%%%%%%%%%%%%%%%%%%%%%%%%%%%%%%%%%%%%%%%%%%%%%%%%%%%%%%%%

The hydrodynamic model requires {\em initial conditions} at the 
earliest time at which the assumption of local thermal equilibrium is 
applicable, and a {\em ``freeze-out prescription''} at the end when the
system becomes too dilute to maintain local thermal equilibrium. Both
are described in detail elsewhere \cite{RANP04}. Different approaches
to freeze-out invoke either the Cooper-Frye algorithm \cite{Cooper:1974mv} 
(used by us), in which chemical freeze-out of the hadron
abundances at $\Tc$ \cite{BMMRS01} must be implemented by hand by
introducing non-equilibrium chemical potentials below $\Tc$ 
\cite{Hirano02,Rapp02,KR03}, or a hybrid approach 
\cite{Bass:2000ib,Teaney:2001cw} that switches from a hydrodynamic 
description to a microscopic hadron cascade at the quark-hadron 
transition, letting the cascade handle the chemical and thermal
freeze-out kinetics. While the radial flow patterns from the two
freeze-out procedures don't differ much, the elliptic flow can be 
quite different if the spatial deformation of the fireball is still
significant during the hadronic stage of the expansion, as I will 
discuss in Sec.~\ref{sec5b}.

We have solved the relativistic equations for ideal hydrodynamics 
under the simplifying assumption of boost-invariant longitudinal 
expansion (see \cite{KSH00,Kolb:2003dz} for details). This is adequate
near midrapidity (the region which most RHIC experiments cover best),
but not sufficient to describe the rapidity distribution of emitted
hadrons and of their transverse flow pattern which require a 
(3+1)-dimensional hydrodynamic code such as the one by Hirano 
\cite{Hirano02}. 

The initial and final conditions for the hydrodynamic evolution are fixed
by fitting the pion and proton spectra at midrapidity in {\em central} 
($b\eq0$) collisions; additionally, we use the centrality dependence 
of the total charged multiplicity $dN_{\rm ch}/dy$. I stress that this 
is the {\em only} information used from $b{\,\ne\,}0$ collisions, and
it is necessary to fix the ratio of soft to hard collision processes
in the initial entropy production. An upper limit for the initial 
ther\-malization time $\tau_0{\,\leq\,}0.6$\,fm/$c$, the initial entropy 
density $s_0\eq110$\,fm$^{-3}$ in the fireball center (corresponding
to an initial maximal energy density $e_0{\,\approx\,}30$\,GeV/fm$^3$ and 
an initial central fireball temperature 
$T_0{\,\approx\,}360\,{\rm MeV}{\,\approx\,}2\,\Tc$), the baryon to
entropy ratio, and the freeze-out energy density 
$e_{\rm dec}\eq0.075$\,GeV/fm$^3$ are all fixed from the $b\eq0$ 
pion and proton spectra (these numbers refer to 200\,$A$\,GeV Au+Au 
collisions at RHIC \cite{KR03}). The initial 
entropy density profile is calculated for all $b$ from the collision 
geometry, using the Glauber model with soft/hard ratio as fixed above.
Except for pions and protons, all other hadron spectra in $b\eq0$
collisions and all spectra for $b{\,\ne\,}0$ collisions (including 
all flow anisotropies such as $v_2$ which vanish at $b\eq0$) are then 
parameter-free predictions of the model. Note that all calculated 
hadron spectra include feeddown from decays of unstable hadron 
resonances.

\vspace*{-3mm}
%%%%%%%%%%%%%%%%%%%%%%%%%%%%%%%%%%%%%%%%%%%%%%%%%%%%%%%%%%%%%%%%% 
\section{Successes of ideal fluid dynamics at RHIC}
\label{sec4}
%%%%%%%%%%%%%%%%%%%%%%%%%%%%%%%%%%%%%%%%%%%%%%%%%%%%%%%%%%%%%%%%% 
\subsection{Hadron momentum spectra and radial flow}
\label{sec4a}
%%%%%%%%%%%%%%%%%%%%%%%%%%%%%%%%%%%%%%%%%%%%%%%%%%%%%%%%%%%%%%%%% 

\begin{figure}[hbt]
%\vspace*{-5mm}
\bce
 \includegraphics[width=0.49\linewidth,height=49.7mm]{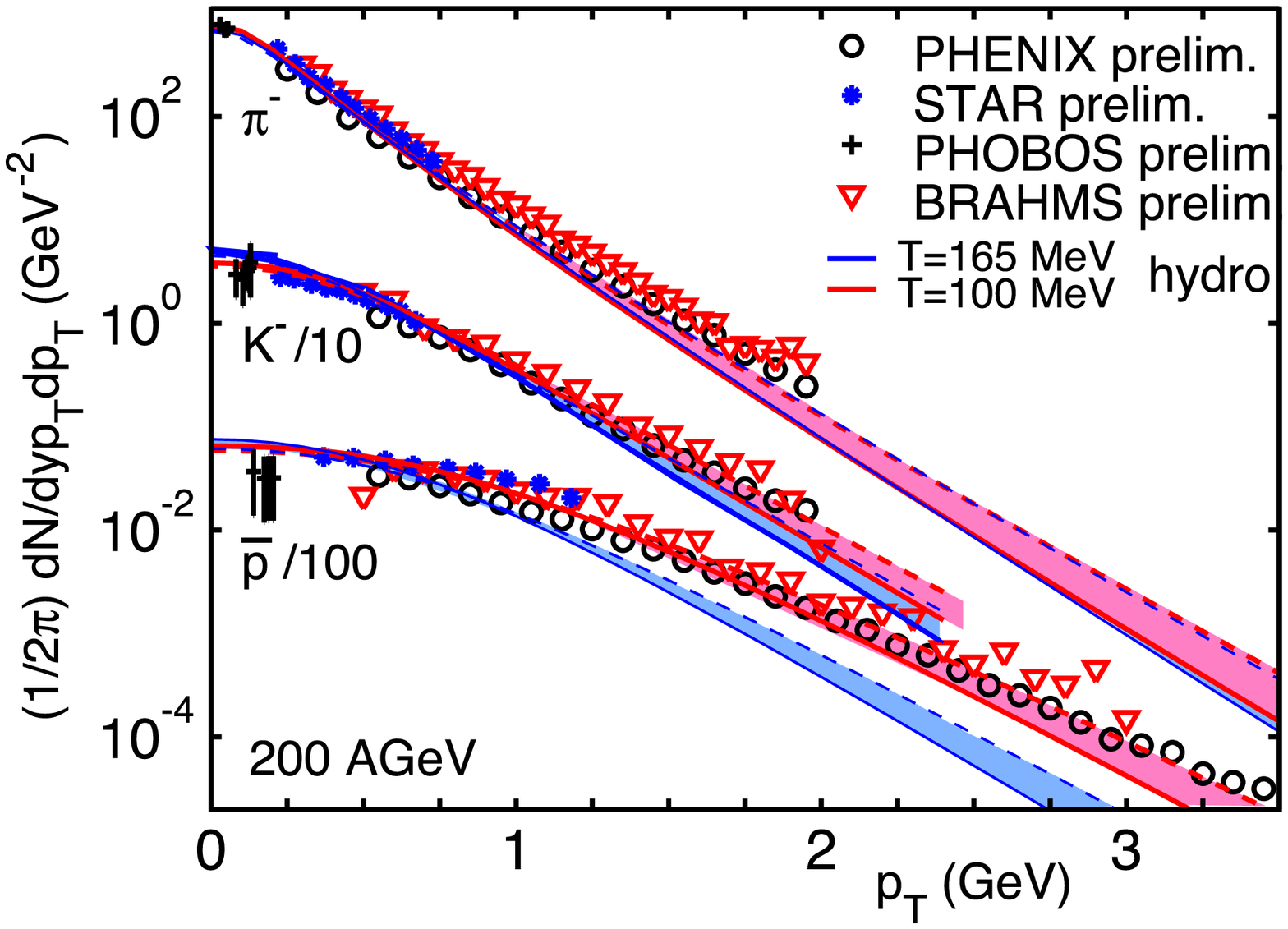}
 \includegraphics[bb=8 30 513 405,width=0.49\linewidth,height=50mm]%
 {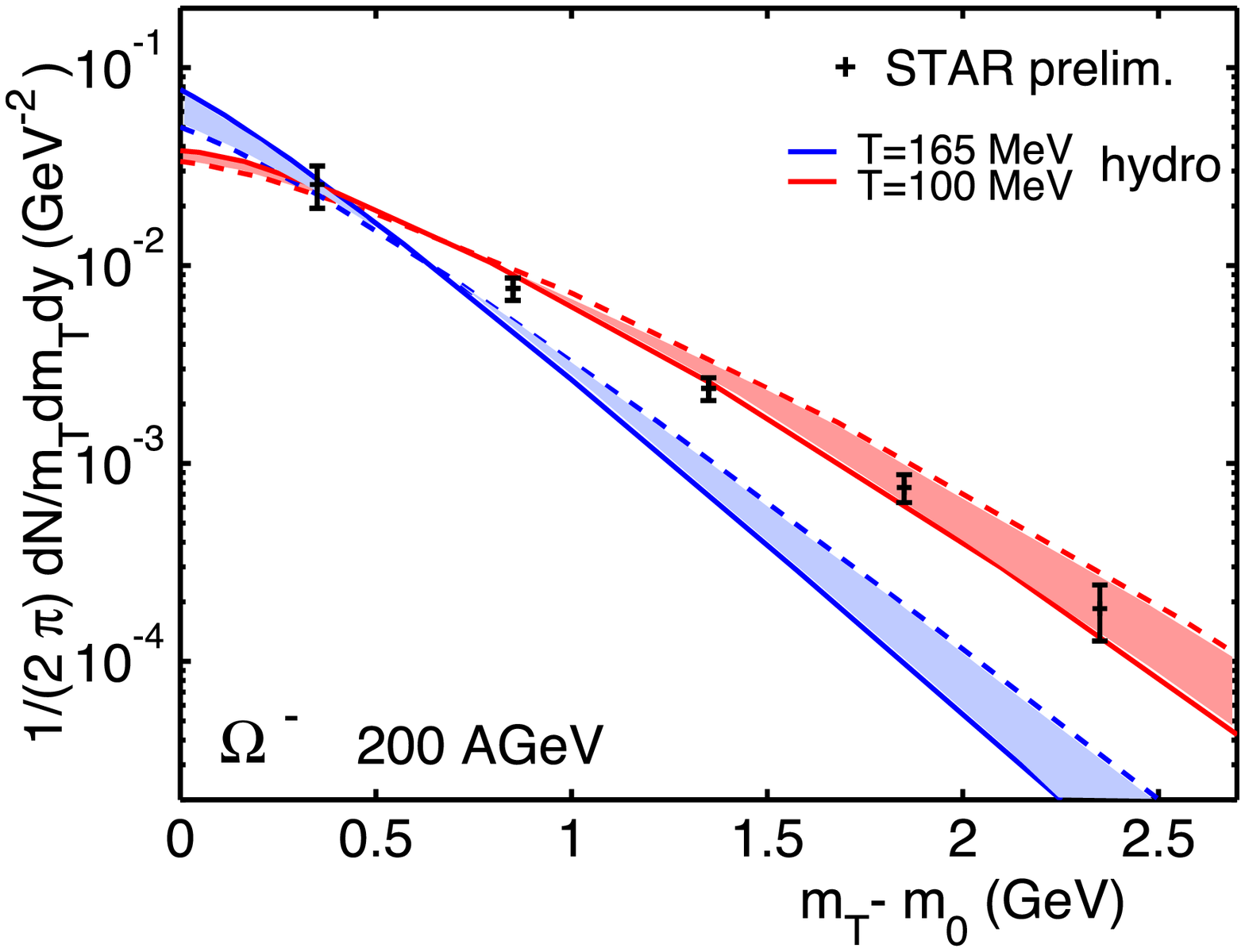}
\ece
\vspace*{-5mm}
\caption{\label{F2} \small
Negative pion, kaon, antiproton, and $\Omega$ spectra from 
central Au+Au collisions at $\sqrt{s}\eq200\,A$\,GeV measured at
RHIC \cite{spec200}. The curves show hydrodynamical calculations
\cite{KR03} (see text).}
\vspace*{-5mm}
\end{figure}

Figure~\ref{F2} shows the single particle $p_\perp$-spectra for 
pions, kaons and antiprotons (left panel) as well as $\Omega$ baryons 
(right panel) measured in Au+Au collisions at RHIC, together with 
hydrodynamical results \cite{KR03}. In order to illustrate the effect 
of additional radial flow generated in the late hadronic stage below 
$T_{\rm cr}$, two sets of curves are shown: the lower (blue) bands 
correspond to kinetic decoupling at $T_{\rm cr}\eq165$\,MeV, whereas 
the upper (red) bands assume decoupling at $T_{\rm dec}\eq100$\,MeV. 
The width of the bands indicates the sensitivity of the calculated 
%
%%%%%%%%%%%%%%%%%%%%%%%%% Fig. 2 %%%%%%%%%%%%%%%%%%%%%%%%%%%%%%%%%%%%%%%%%%%%
%%%%%%%%%%%%%%%%%%%%%%%%%%%%%%%%%%%%%%%%%%%%%%%%%%%%%%%%%%%%%%%%%%%%%%%%%%%%%%%
%
spectra to an initial transverse flow of the fireball already at the 
time of thermalization \cite{KR03}. In the hydrodynamic simulation it 
takes about 9-10\,fm/$c$ until most of the fireball becomes 
sufficiently dilute to convert to hadronic matter and another 
7-8\,fm/$c$ to fully decouple \cite{KSH00}. Figure~\ref{F2} shows 
that by the time of hadronization the dynamics has not yet generated 
enough radial flow to reproduce the measured $\bar p$ and $\Omega$ 
spectra; these heavy hadrons, which are particularly sensitive to 
radial flow, require the additional collective ``push'' created by 
resonant quasi-elastic interactions during the fairly long-lived 
hadronic rescattering stage. The flattening of the $\bar p$
spectra by radial flow provides a natural explanation for 
the (initially puzzling) experimental observation that for 
$p_\perp{\,>\,}2$\,GeV/$c$ antiprotons become more abundant than
pions \cite{Kolb:2003dz}.

As shown elsewhere (see Fig.~1 in \cite{Heinz:2002un}), the model 
describes these and all other hadron spectra not only in central, but 
also in {\it peripheral} collisions, up to impact parameters of about 
10\,fm, and with similar quality. No additional parameters enter at 
non-zero impact parameter.

\vspace*{-3mm}
%%%%%%%%%%%%%%%%%%%%%%%%%%%%%%%%%%%%%%%%%%%%%%%%%%%%%%%%%%%%%%%%% 
\subsection{Elliptic flow}
\label{sec4b}
%%%%%%%%%%%%%%%%%%%%%%%%%%%%%%%%%%%%%%%%%%%%%%%%%%%%%%%%%%%%%%%%% 

\begin{figure}[htb]
\vspace*{-2mm}
\bce
\includegraphics*[bb=61 205 568 594,width=0.44\linewidth,height=49mm]%
                 {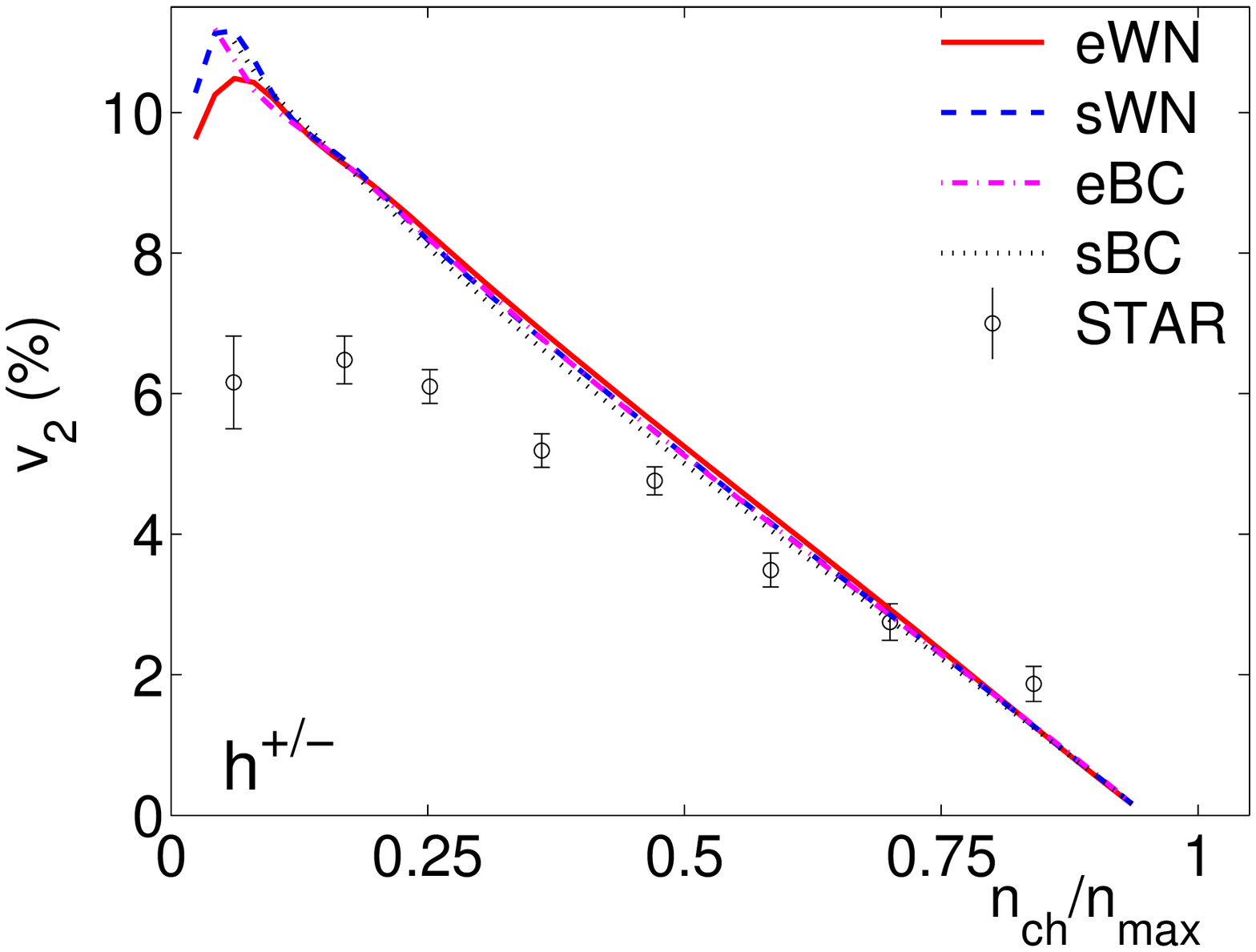}
\includegraphics*[bb=0 10 567 470,width=0.54\linewidth,height=52mm]%
                 {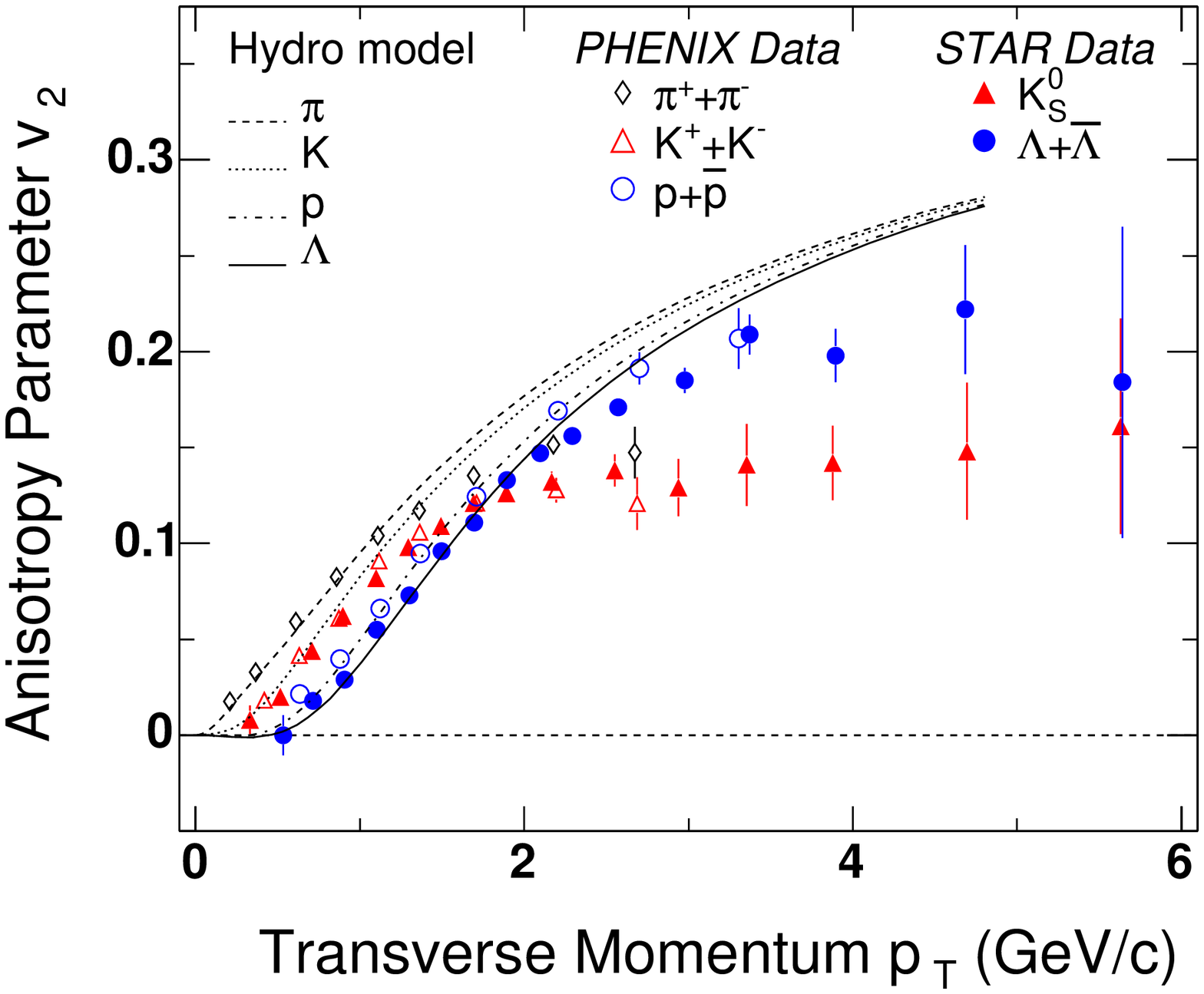}
\ece
\vspace*{-5mm}
\caption{\label{F3} \small
Left: $p_\perp$-averaged elliptic flow for all charged hadrons from 
130\,$A$\,GeV Au+Au collisions, as a function of collision 
centrality ($n_{\rm ch}$ is the charged multiplicity at $y\eq0$).
The curves are hydrodynamic calculations with different choices for 
the initial energy density profile (see \cite{KHHET}).
Right: Differential elliptic flow $v_2(p_\perp)$ for identified 
hadrons from minimum bias Au+Au collisions at 200\,$A$\,GeV 
\cite{Ackermann:2001tr,PHENIXv2,Sorensen:2003kp}, together with 
hydrodynamic curves from \cite{Huovinen:2001cy}.
\vspace*{-3mm}
}
\end{figure}

Figure~\ref{F3} shows the predictions for the elliptic flow coefficient 
$v_2$ from Au+Au collisions at RHIC, together with the data 
\cite{Ackermann:2001tr,PHENIXv2}. For impact parameters 
$b{\,\leq\,}7$\,fm ($n_{\rm ch}/n_{\rm max}{\,\geq\,}0.5$)
%
%%%%%%%%%%%%%%%%%%%%%%%%% Fig. 3 %%%%%%%%%%%%%%%%%%%%%%%%%%%%%%%%%%%%%%%%%%%%
%%%%%%%%%%%%%%%%%%%%%%%%%%%%%%%%%%%%%%%%%%%%%%%%%%%%%%%%%%%%%%%%%%%%%%%%%%%%%%%
%
and transverse momenta $p_\perp{\,\lapp\,}1.5$\,GeV/$c$ the data
are seen (left panel of Fig.~\ref{F3}) to exhaust the upper limit for 
$v_2$ obtained from the hydrodynamic calculations. For larger impact 
parameters $b{\,>\,}7$\,fm the $p_\perp$-averaged elliptic flow $v_2$ 
increasingly lags behind the hydrodynamic prediction; this will be 
discussed in detail in Sec.~\ref{sec5b}. As a function of $\pt$ (right 
panel of Fig.~\ref{F3}) the elliptic flow of all hadrons measured so far is 
very well described by hydrodynamics, for $\pt{\,\lapp\,}1.5-2$\,GeV/$c$. 
In particular the hydrodynamically predicted {\em mass splitting} of 
$v_2$ at low $p_\perp$ is perfectly reproduced by the data. This mass 
splitting depends on the EOS \cite{Huovinen:2001cy}, and the EOS 
including a quark-hadron phase transition used here describes
the data better than one without phase transition (see Fig.~2 
in \cite{Heinz:2002un}). Ideal fluid dynamics with a QGP EOS
thus gives an excellent and very detailed description of 
{\em all} hadron spectra below $p_\perp\eq1.5$\,GeV/$c$. Since this 
$\pt$-range includes more than 99\% of all produced hadrons, it is 
fair to say that {\em the bulk of the fireball matter formed in Au+Au 
collisions at RHIC behaves very much like a perfect fluid.}

\vspace*{-3mm}
%%%%%%%%%%%%%%%%%%%%%%%%%%%%%%%%%%%%%%%%%%%%%%%%%%%%%%%%%%%%%%%%% 
\subsection{Final source eccentricity in coordinate space}
\label{sec4c}
%%%%%%%%%%%%%%%%%%%%%%%%%%%%%%%%%%%%%%%%%%%%%%%%%%%%%%%%%%%%%%%%% 

While spectra and elliptic flow reflect the {\em momentum} structure of
the hadron emitting source, its {\em spatial} deformation can be tested 
with 2-pion Hanbury Brown-Twiss (HBT) correlations measured as a 
function of the azimuthal emission angle 
\cite{HK02,RL04}. Even though the initial spatial deformation of the 
reaction zone in non-central Au+Au collisions at RHIC is finally 
completely gone, many pions are already emitted from earlier times 
when the spatial deformation is still significant. For Au+Au collisions
at $b\eq7$\,fm, the spatial eccentricity of the time-integrated 
hydrodynamic pion emission function is $\langle\varepsilon_x\rangle\eq0.14$ 
(or 56\% of its initial value $\varepsilon_x(\tau_0)\eq0.25$)
\cite{HK02}. Using azimuthally sensitive pion HBT measurements, the 
STAR Collaboration has measured in the corresponding impact parameter 
bin \cite{STARasHBT} $\langle\varepsilon_x\rangle\eq0.11\pm0.035$ 
(or $45\pm15\%$ of the initial deformation). This agreement can be 
counted as another success for the 
hydrodynamic model.
  
\vspace*{-3mm}
%%%%%%%%%%%%%%%%%%%%%%%%%%%%%%%%%%%%%%%%%%%%%%%%%%%%%%%%%%%%%%%%%%%%%%%
\section{Viscous effects at RHIC}
\label{sec5}
%%%%%%%%%%%%%%%%%%%%%%%%%%%%%%%%%%%%%%%%%%%%%%%%%%%%%%%%%%%%%%%%%%%%%%%
\subsection{QGP viscosity}
\label{sec5a}
%%%%%%%%%%%%%%%%%%%%%%%%%%%%%%%%%%%%%%%%%%%%%%%%%%%%%%%%%%%%%%%%%%%%%%%

As evident in the right panel of Fig.~\ref{F3}, the hydrodynamic 
prediction for $v_2(\pt)$ gradually breaks down above 
$\pt{\,\gapp\,}1.5$\,GeV/$c$ for mesons and above 
$\pt{\,\gapp\,2.2}$\,GeV/$c$ for baryons. The empirical fact 
\cite{Sorensen:2003kp} that both the $\pt$-values, where this 
break from hydrodynamics sets in, and the saturation values 
for $v_2$ at high $\pt$ for baryons and mesons are always 
({\em i.e. for all collision cen\-tralities!}) related by 3\,:\,2 (i.e. by 
their ratios of valence quark numbers), independent of their masses, 
tells its own interesting story (see e.g. \cite{coal}): It strongly
suggests that in this $\pt$ region hadrons are formed
by coalescence of color-deconfined quarks, and that the elliptic
flow is of partonic origin (i.e. generated before hadronization),
with a $\pt$-shape that follows hydrodynamics at low $\pt$ up to
about 750\,MeV and then gradually breaks away \cite{coal}.  

\begin{figure}[htb]
\vspace*{-10mm}
\bce
\includegraphics*[width=0.49\linewidth,height=52mm]{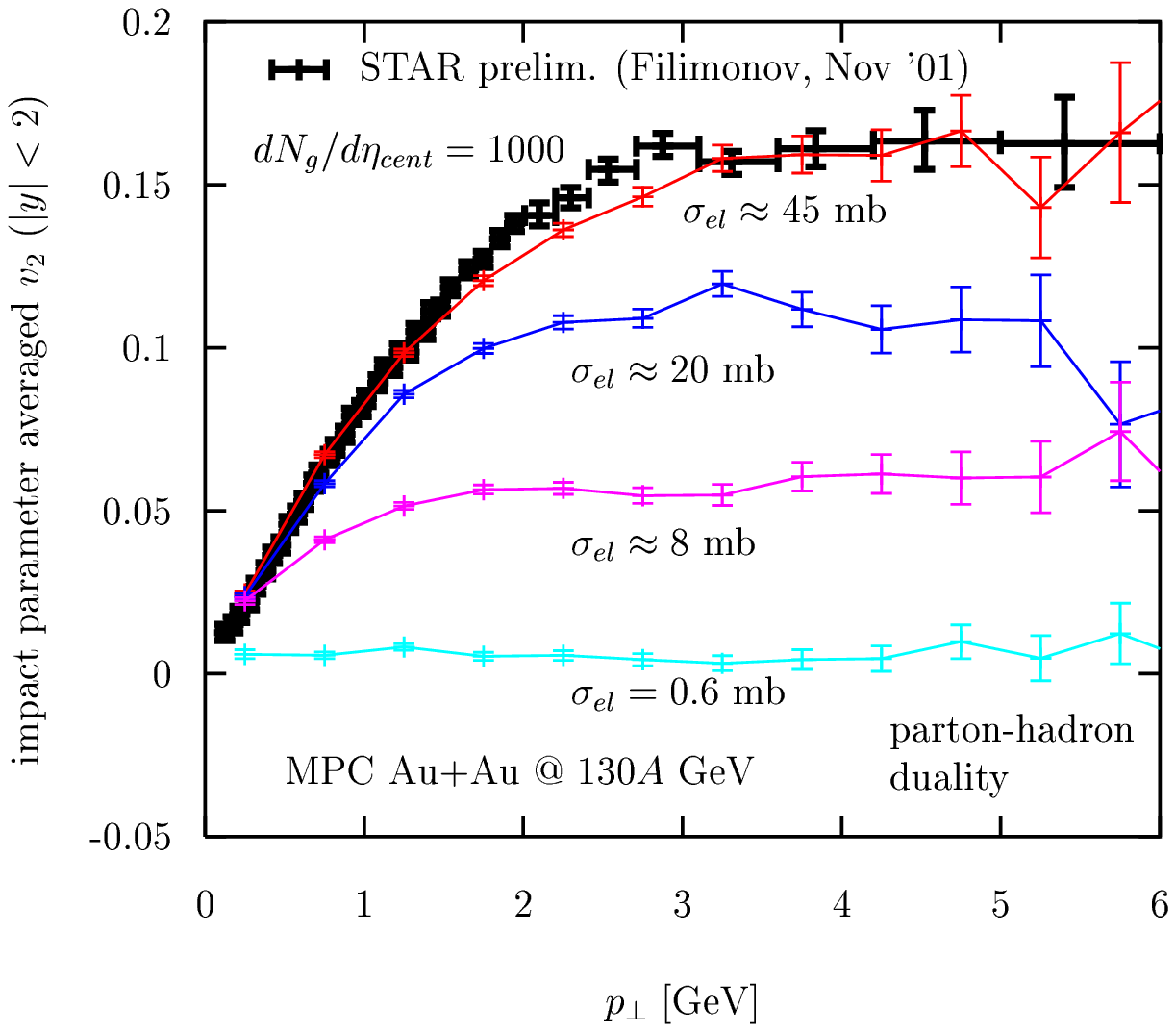}
\includegraphics*[bb=0 -43 568 594,width=0.49\linewidth,height=61mm,clip=]%
{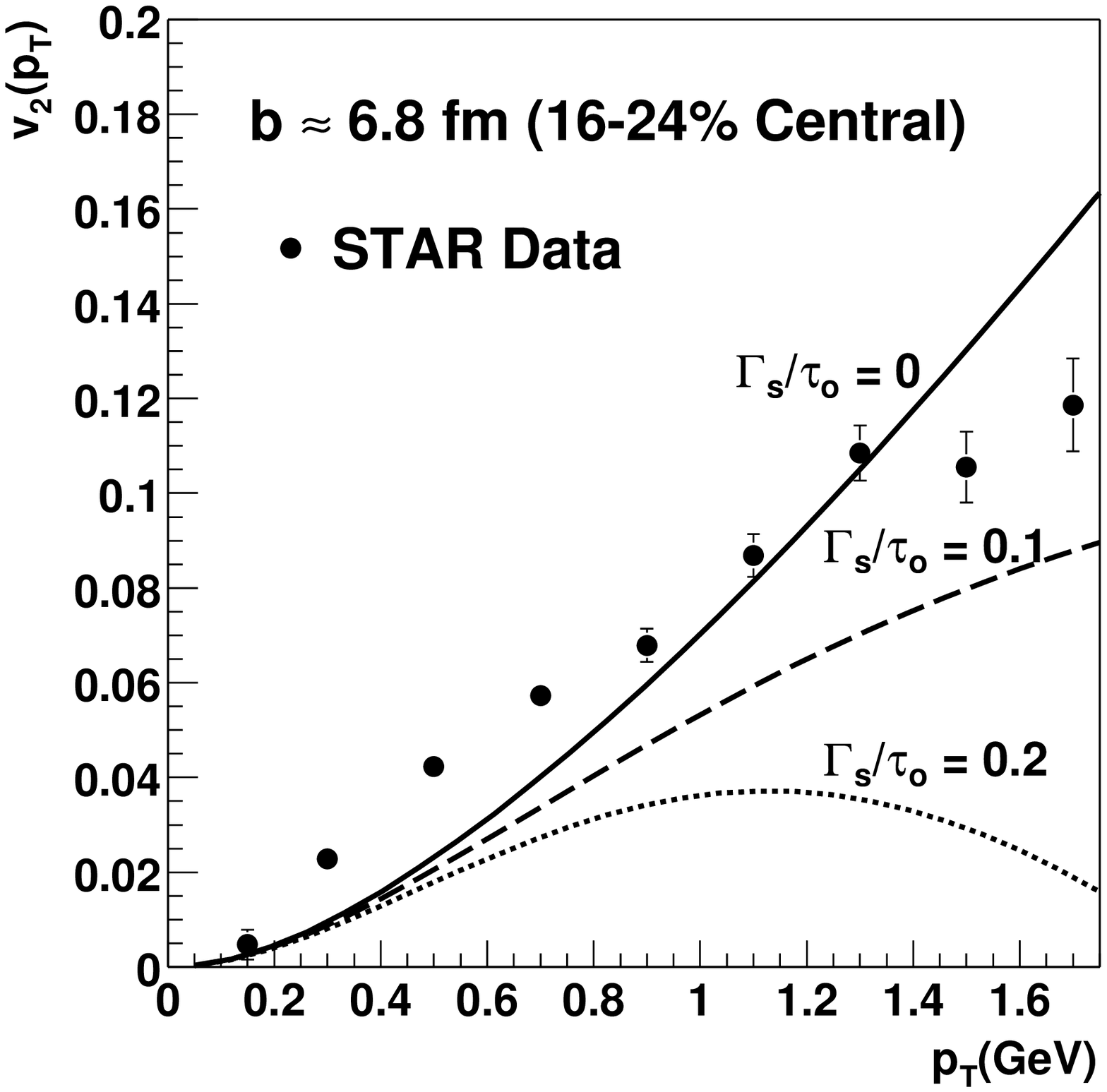}
\ece
\vspace*{-7mm}
\caption{\label{F4}\small 
Left: Elliptic flow from a parton cascade \cite{Molnar:2001ux}, compared
with STAR data, for different parton-parton scattering cross sections.
Larger cross sections lead to smaller mean free paths. Right: Perturbative
effects of shear viscosity on the elliptic flow $v_2(p_\perp)$
\cite{T03} (see text).
}
\vspace*{-4mm}
\end{figure}

Since $v_2(p_\perp)$ is a measure for the relatively small differences 
between the in-plane and out-of-plane slopes of the $\pt$ spectra, it 
is more sensitive to deviations from ideal fluid dynamic 
behaviour than the angle-averaged slopes. Two model studies 
\cite{Heinz:2002rs,T03} showed that $v_2$ reacts particularly strongly 
to shear viscosity. As the mean free path of the plasma constituents 
(and thus the fluid's viscosity) goes to zero, $v_2$ approaches the 
ideal fluid limit from below \cite{Molnar:2001ux} (see Fig.\,\ref{F4}a). 
At higher transverse momenta it does so more slowly than at low 
$p_\perp$ \cite{Molnar:2001ux}, approaching a constant saturation 
value at high $\pt$. The increasing deviation from the
%
%%%%%%%%%%%%%%%%%%%%%%%%% Fig. 4 %%%%%%%%%%%%%%%%%%%%%%%%%%%%%%%%%%%%%%%%%%%%
%%%%%%%%%%%%%%%%%%%%%%%%%%%%%%%%%%%%%%%%%%%%%%%%%%%%%%%%%%%%%%%%%%%%%%%%%%%%%%%
%
ideal fluid limit for growing $\pt$ is qualitatively consistent with 
the expected influence of shear viscosity: Teaney \cite{T03} showed 
that the lowest order viscous corrections to the local equilibrium 
distribution increase quadratically with $\pt$ so that $v_2$ remains 
increasingly below the ideal fluid limit as $p_\perp$ grows (see 
Fig.\,\ref{F3}b). From Fig.\,\ref{F4}b Teaney concluded that at RHIC 
the normalized sound attenuation length 
$\frac{\Gamma_s}{\tau} = \frac{4}{3T\tau}\frac{\eta}{s}$ (where $\eta$ is 
the shear viscosity, $T$ the temperature and $s$ the entropy density) 
cannot be much larger than about 0.1. This puts a stringent limit on 
the dimensionless ratio $\eta/s$, bringing it close to the recently
conjectured absolute lower limit for the viscosity of
$\eta/s\eq\hbar/(4\pi)$ \cite{son}. This would make the quark-gluon
plasma the most ideal fluid ever observed \cite{son}.

These arguments show that deviations from ideal fluid dynamics at
high $p_\perp$ must be expected, and that they can be large even
for fluids with very low viscosity. At which $p_\perp$ non-ideal 
effects begin to become visible in $v_2(p_\perp)$ can be taken 
as a measure for the fluid's viscosity. To answer the {\em quantitative}
question what the RHIC data on partonic elliptic flow and its 
increasing deviation from ideal fluid behaviour above 
$\pt{\,\gapp\,}750$\,MeV/$c$ imply for the {\em value} of the QGP 
shear viscosity $\eta$ requires a numerical algorithm for solving 
viscous relativistic hydrodynamics (see \cite{asis} for an update).

\vspace*{-3mm}
%%%%%%%%%%%%%%%%%%%%%%%%%%%%%%%%%%%%%%%%%%%%%%%%%%%%%%%%%%%%%%%%%%%%%%%
\subsection{Viscosity of the hadron resonance gas}
\label{sec5b}
%%%%%%%%%%%%%%%%%%%%%%%%%%%%%%%%%%%%%%%%%%%%%%%%%%%%%%%%%%%%%%%%%%%%%%%

\begin{figure}[htb]
\vspace*{-5mm}
\begin{minipage}[t]{0.5\linewidth}
\includegraphics*[bb=0 0 567 470,width=\linewidth,height=55mm]{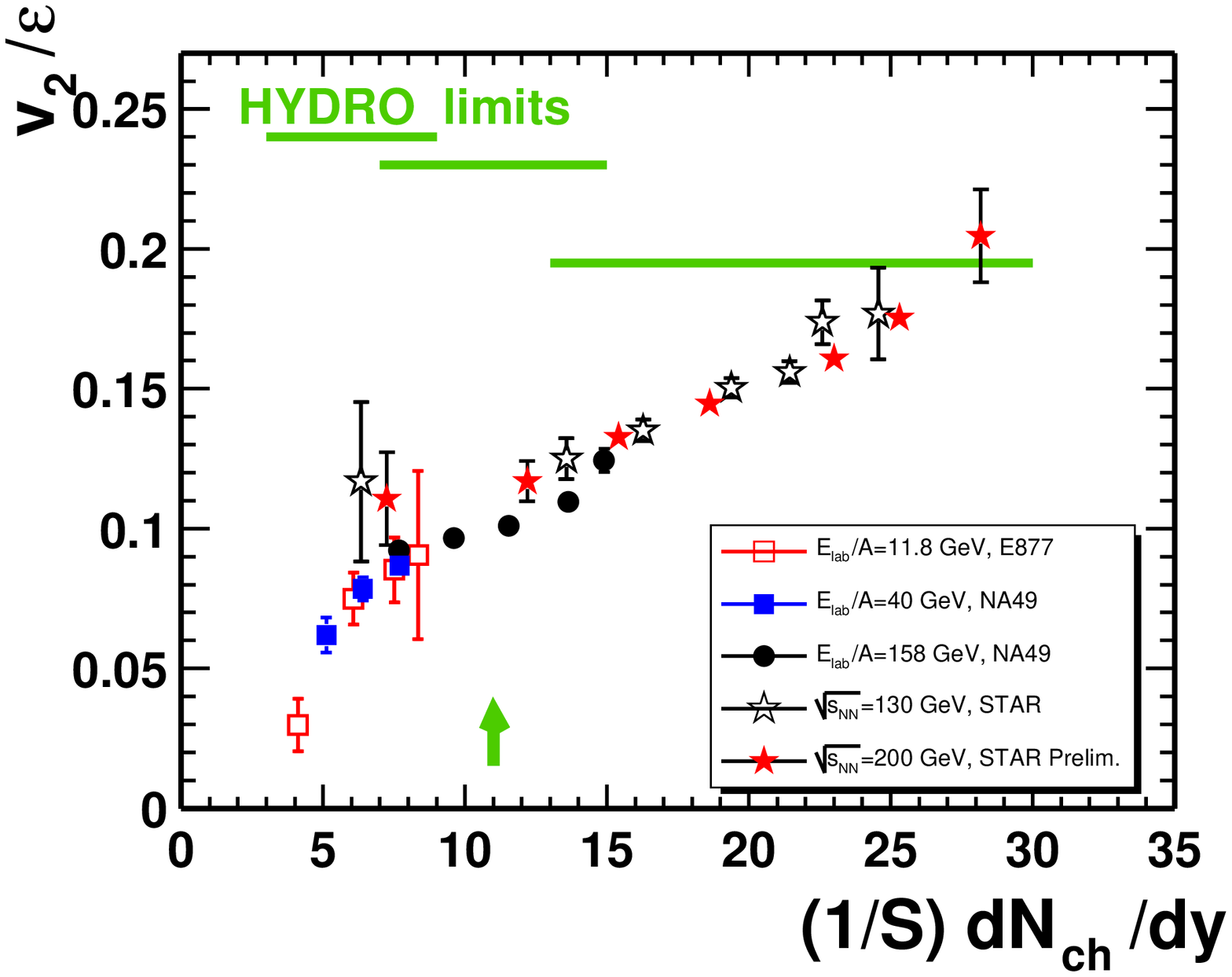}
\end{minipage}
\begin{minipage}[b]{0.48\linewidth}
\includegraphics*[bb=0 -60 568 490,width=\linewidth,height=44.4mm]{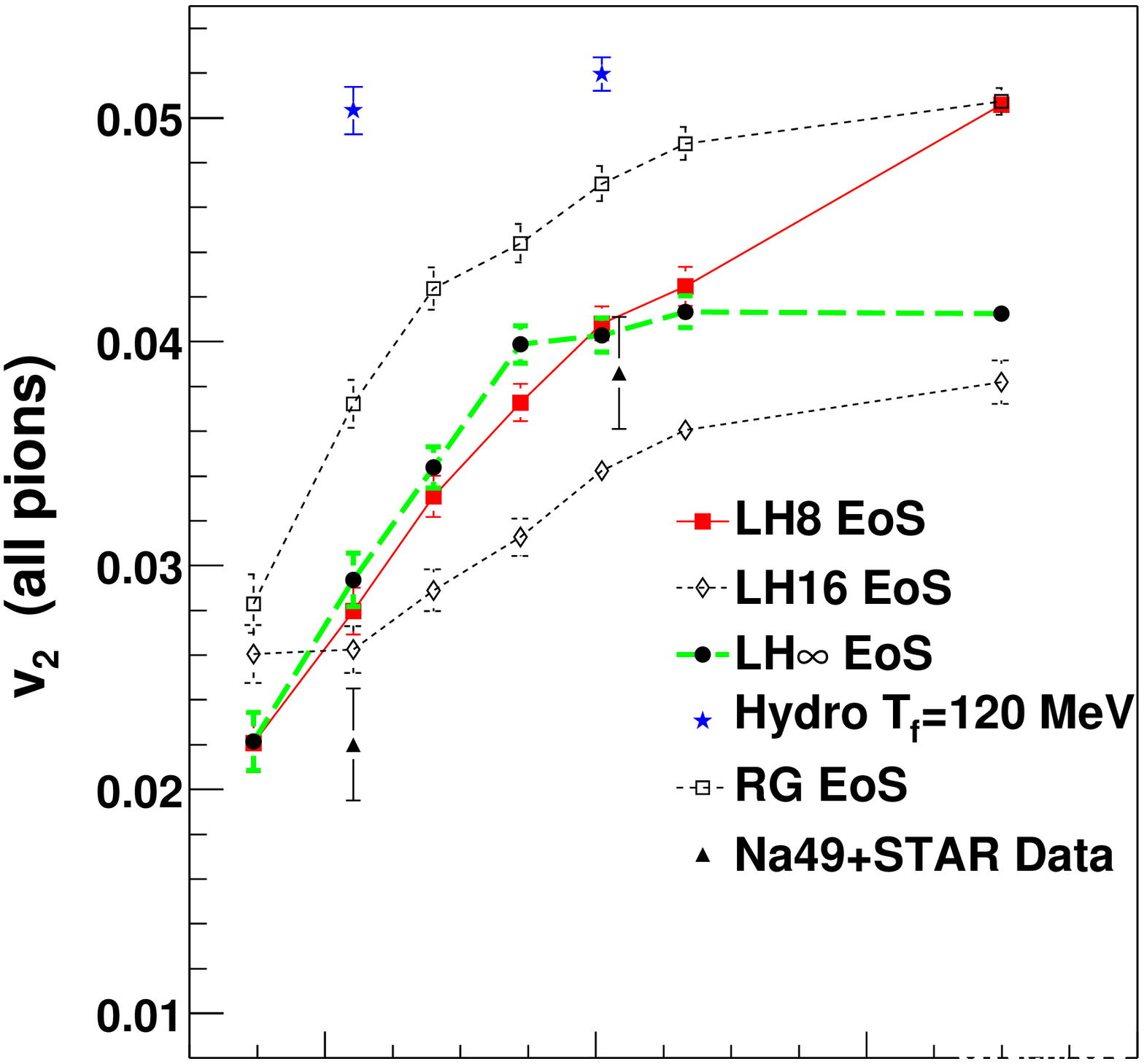}\\[-6mm]
\includegraphics*[bb=0 0 567 68,width=\linewidth,height=8mm]{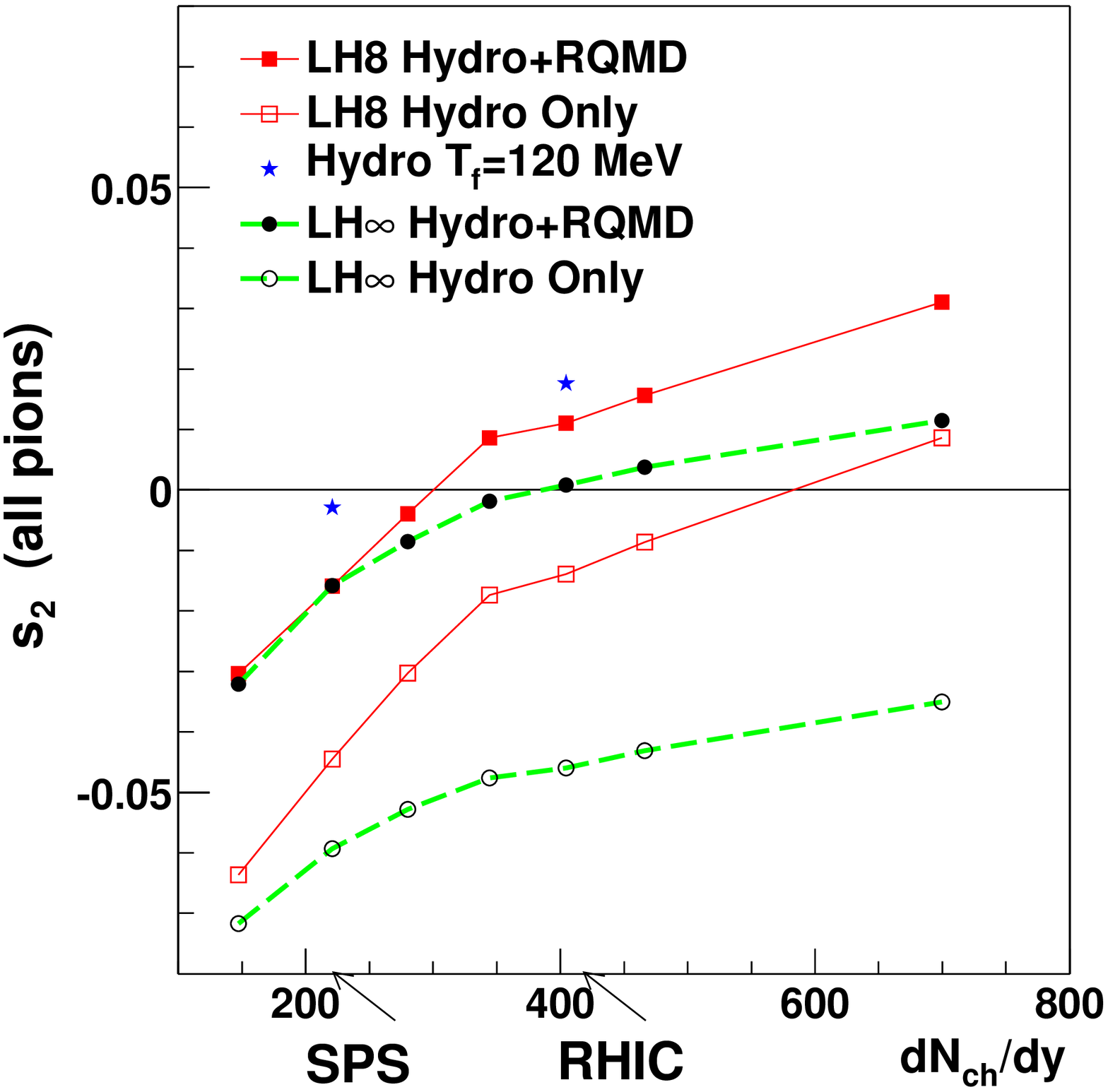}\\[-2.3mm]
\end{minipage}
\vspace*{-5mm}
\caption{\label{F5} \small
Left: Scaled elliptic flow $v_2/\varepsilon$ (where $\varepsilon$ 
denotes the initial spatial eccentricity) vs. the charged multiplicity 
density per unit initial transverse overlap area $S$ \cite{NA49v2PRC}.
Right: Elliptic flow $v_2$ for minimum bias Au+Au collisions
at various collision energies, parametrized by the final charged
multiplicity density at midrapidity \cite{Teaney:2001cw}. See text
for discussion.
}
\vspace*{-2mm}
\end{figure}

\begin{figure}[hb]
\vspace*{-3mm}
\bce
\includegraphics*[bb=49 165 579 616,width=0.53\linewidth,height=5.1cm]%
                 {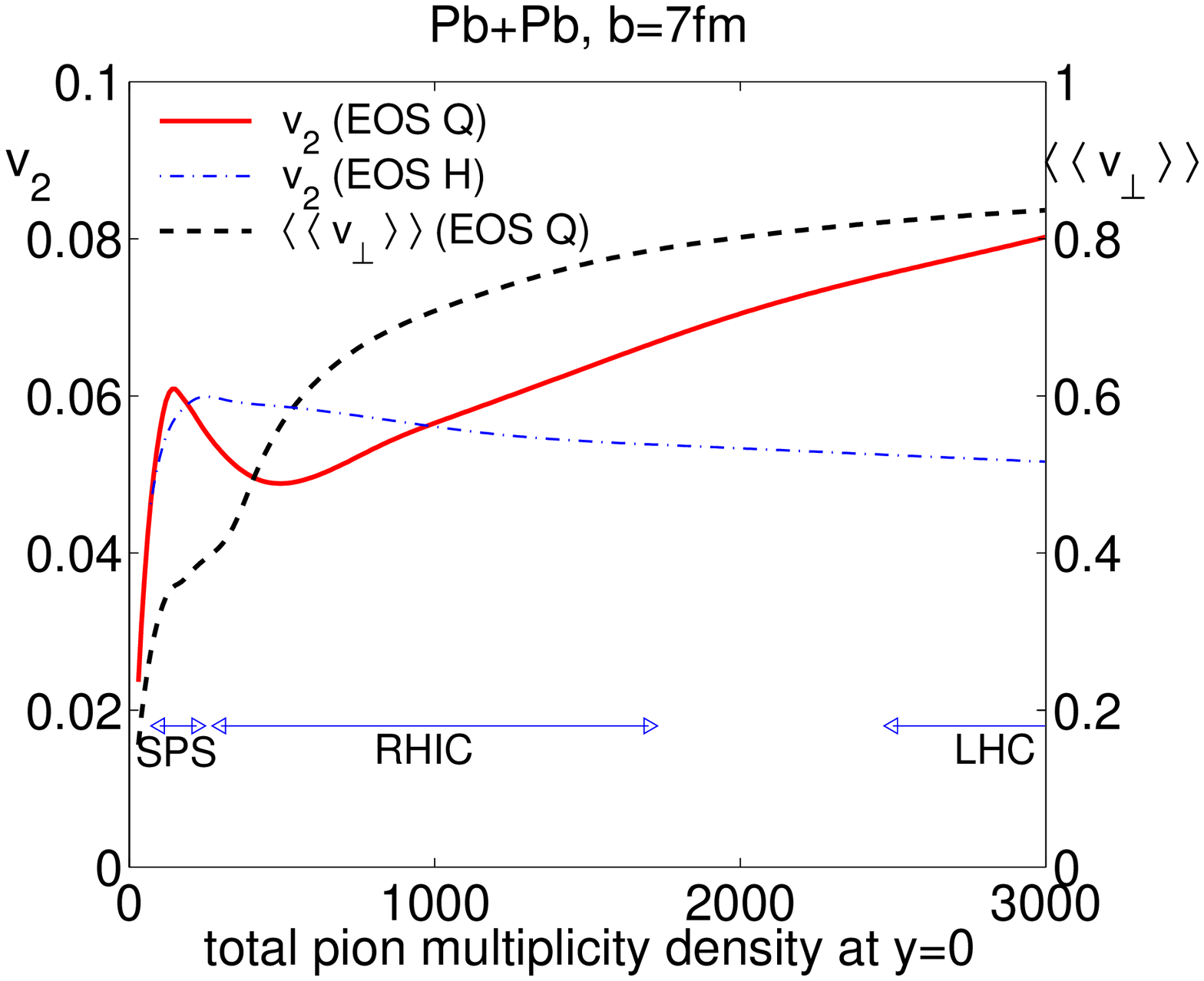}
\includegraphics*[bb=116 396 502 683,width=0.45\linewidth,height=4.7cm]%
                 {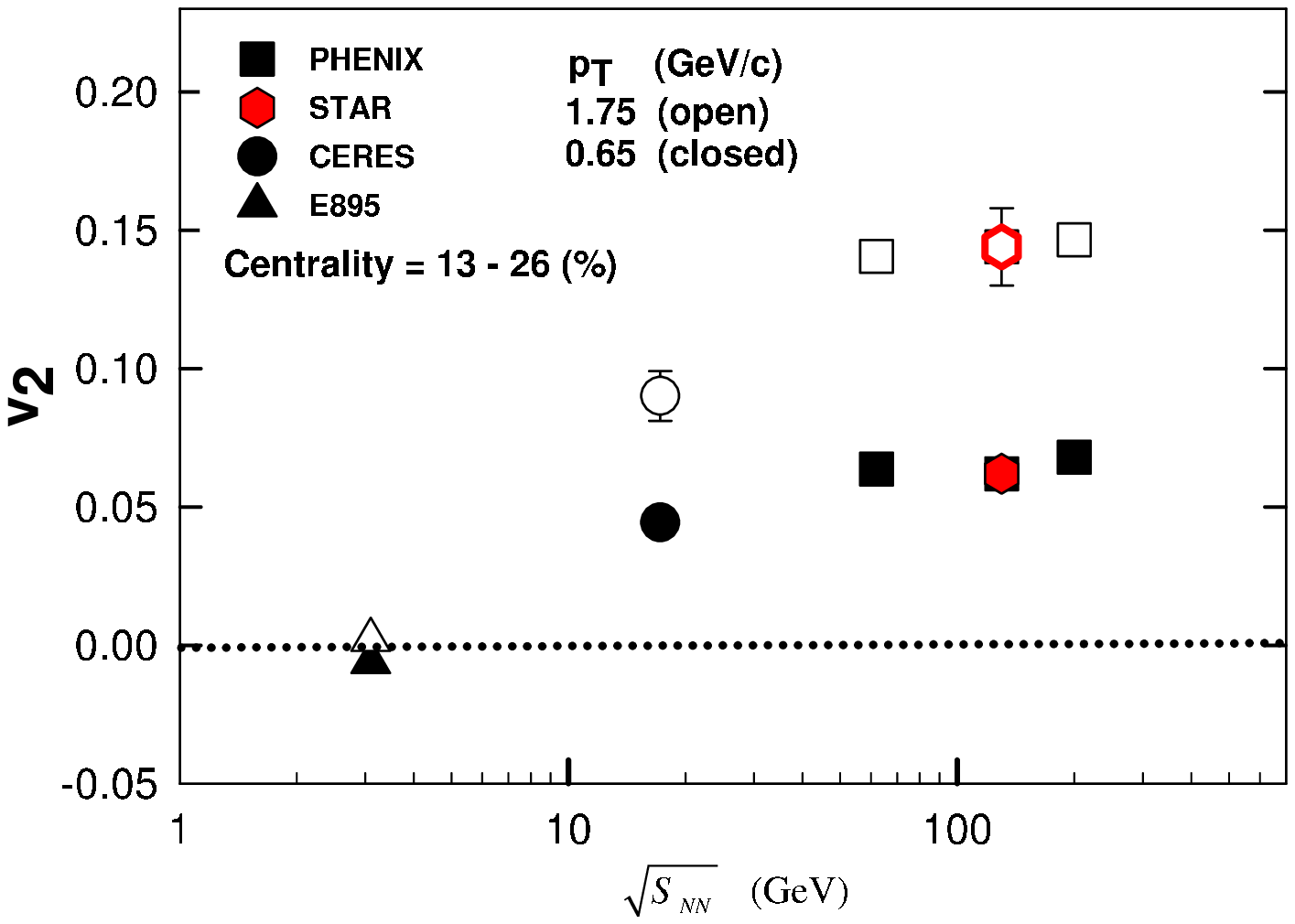}
\ece
\vspace*{-5mm}
\caption{\label{F6}\small
Left: Excitation function of radial and $\pt$-integrated elliptic flow
for $b\eq7$\,fm Pb+Pb or Au+Au collisions \cite{KSH00}. The horizontal
axis gives $dN_\pi/dy(b{=}7\,{\rm fm})$, and the horizontal arrows
indicate how $dN_\pi/dy$ was expected to correspond to the beam energy 
ranges covered by SPS, RHIC and LHC before the RHIC data were available. 
Right: Elliptic flow at fixed $\pt\eq0.65$ and 1.75\,GeV/$c$ from A+A 
collisions with $A{\,\approx\,}200$ at a variety of beam energies 
\cite{PHENv2rs}. 
}
\vspace*{-3mm}
\end{figure}

Ideal fluid dynamics also fails to describe the elliptic flow $v_2$ in 
more peripheral Au+Au collisions at RHIC and in central and peripheral 
collisions at lower energies (see Fig.~\ref{F5}a), as well as 
at forward rapidities in minimum bias Au+Au collisions at RHIC 
\cite{Hirano:2001eu}. Whereas hydrodynamics predicts a non-monotonic 
beam energy dependence of $v_2$ (Fig.~\ref{F6}a \cite{KSH00}), with 
largest values at upper AGS and lower SPS energies, somewhat lower 
values at RHIC and again larger values at the LHC, the data seem to 
increase monotonically with $\sqrt{s}$.

%
%%%%%%%%%%%%%%%%%%%%%%%%% Fig. 5 %%%%%%%%%%%%%%%%%%%%%%%%%%%%%%%%%%%%%%%%%%%%
%%%%%%%%%%%%%%%%%%%%%%%%%%%%%%%%%%%%%%%%%%%%%%%%%%%%%%%%%%%%%%%%%%%%%%%%%%%%%%%
%
Such a monotonic rise is consistent with ``hybrid'' calculations by Teaney 
\cite{Teaney:2001cw} (Fig.~\ref{F5}b) where the fireball undergoes ideal 
fluid dynamic evolution only while in the QGP, followed by hadronic 
{\em kinetic} evolution using RQMD after hadronization. Figure~\ref{F5}b 
shows several curves corresponding to different equations of state during 
the hydrodynamic evolution (see \cite{Teaney:2001cw}), with LH8 
being closest to the data. The difference 
between the points labelled LH8 and the hydrodynamic values at the
top of the figure is due to the different evolution during the 
late hadronic stage. Obviously, at lower collision 
energies and for impact parameters $b{\,\sim\,}7$\,fm (which dominate 
$v_2$ in minimum bias collisions), ideal fluid dynamics 
continues to build additional elliptic flow during the hadronic stage, 
but RQMD does not. The initial energy densities are smaller than at RHIC
and the fireball does not spend enough time in the QGP phase for the 
spatial eccentricity $\varepsilon$ to fully disappear before entering 
the hadron resonance gas phase. Anisotropic pressure gradients thus 
still exist in the hadronic phase, and ideal fluid dynamics reacts to 
them according to the stiffness of the hadron resonance gas EOS 
($p\approx0.15e$). Teaney's calculations \cite{Teaney:2001cw} show 
that RQMD responds to these remaining anisotropies much more weakly, 
building very little if any additional elliptic flow during the 
hadronic phase. The hadron resonance gas is a highly viscous medium, 
unable to maintain local thermal equilibrium. The failure of the 
hydrodynamic model in situations where the initial energy density 
is less than about 10\,GeV/fm$^3$ \cite{Heinz:2004et} is therefore 
likely {\em not} the result of viscous effects in the early 
QGP fluid, but rather caused by the highly viscous late hadronic 
stage which is unable to efficiently respond to any remaining spatial 
fireball eccentricity. Similar arguments hold at forward rapidities at RHIC 
\cite{Heinz:2004et} where the initial energy densities are also 
significantly smaller than at midrapidity while the initial spatial 
eccentricities are similar.

The large hadron gas viscosity spoils one of the clearest experimental
signatures for the quark-hadron phase transition, the predicted \cite{KSH00}
non-monotonic beam energy dependence of $v_2$ which was already
described in Sec.~\ref{sec2} and is shown in Fig.~\ref{F6}a.
As one comes down from infinite beam energy, $v_2$ is predicted 
to first decrease (due to the softening of the EOS in the 
phase transition region) and then recover somewhat in the moderately 
stiff hadron gas phase. The hadron gas viscosity spoils this recovery,
leading to an apparently monotonous decrease of $v_2$ with falling
beam energy (Fig.~\ref{F5}). However, recent PHENIX data from Au+Au 
collisions at $\sqrt{s}\eq62\,A$\,GeV \cite{PHENv2rs} indicate that 
this decrease may not be quite as monotonous as suggested by 
Fig.~\ref{F5}a. Figure~\ref{F6}b shows that, at fixed $\pt$,
the elliptic flow $v_2(\pt)$ is essentially constant over the entire 
energy range explored at RHIC (from 62 to 200 $A$ GeV), decreasing 
only when going further down to SPS and AGS energies. When integrated
over $\pt$, this turns into a monotonous behaviour as in Fig.~\ref{F5}a,
due to the steepening $\pt$ spectra at lower $\sqrt{s}$. While 
Fig.~\ref{F6}b does not confirm the hydrodynamically predicted 
{\em rise} of $v_2$ towards lower $\sqrt{s}$, it may still reflect 
this predicted non-monotonic structure in the elliptic flow excitation 
function, after strong dilution by hadronic viscosity effects.
Obviously, many and more systematic hybrid calculations of the type 
pioneered by Teaney \cite{Teaney:2001cw} are necessary to explore 
to what extent we can eventually prove the existence of a QCD phase 
transition using elliptic flow data in the SPS-RHIC energy domain. 

%
%%%%%%%%%%%%%%%%%%%%%%%%%%%%% Fig.6 %%%%%%%%%%%%%%%%%%%%%%%%%%%%%%%%%%%%%%%
%%%%%%%%%%%%%%%%%%%%%%%%%%%%%%%%%%%%%%%%%%%%%%%%%%%%%%%%%%%%%%%%%%%%%%%%%%%
%
Obviously, it will be important to confirm non-viscous fluid 
behaviour at higher initial energy densities than so far explored, 
by extending Fig.~\ref{F5}a to the right and verifying that 
$v_2/\varepsilon$ settles on the hydrodynamic curve. This can be done 
with Pb+Pb collisions at the LHC, or with full-overlap U+U collisions 
at RHIC \cite{HK05}. In addition, the large spatial source deformations 
achievable in full-overlap side-on-side U+U collisions allow 
for decisive systematic studies of the non-linear path-length dependence 
of QCD radiative energy loss of fast partons. In \cite{HK05} we give
quantitative arguments why a U+U collision program should be seriously
considered at RHIC.    

\vspace*{-3mm}
%%%%%%%%%%%%%%%%%%%%%%%%%%%%%%%%%%%%%%%%%%%%%%%%%%%%%%%%%%%%%%
\section{Conclusions}
%%%%%%%%%%%%%%%%%%%%%%%%%%%%%%%%%%%%%%%%%%%%%%%%%%%%%%%%%%%%%%

The collective flow patterns observed at RHIC provide strong evidence
for fast thermalization at less than 1\,fm/$c$ after impact and at
energy densities more than an order of magnitude above the critical
value for color deconfinement. The thus created thermalized QGP 
is a strongly coupled plasma which behaves like an almost ideal fluid.
These features are first brought out in heavy-ion collisions at RHIC 
because only there the initial energy densities and QGP life times 
are large enough for the ideal fluid character of the QGP to really
manifest itself, in the form of fully saturated hydrodynamic elliptic 
flow, undiluted by late non-equilibrium effects from the highly viscous 
hadron resonance gas which dominates the expansion at lower energies.

We are now ready for a systematic experimental and theoretical program 
to quantitatively extract the EOS, thermalization time and transport 
properties of QGP and hot hadronic matter. This requires more statistics 
and a wider systematic range for soft hadron production data, but more 
importantly a wide range of systematic simulation studies with the 
``hydro-hadro'' hybrid algorithms and, above all, a (3+1)-dimensional
viscous relativistic hydrodynamic code (see \cite{asis} for more on that).   

\vspace*{-3mm}
%%%%%%%%%%%%%%%%%%%%%%%%%%%%%%%%%%%%%%%%%%%%%%%%%%%%%%%%%%%%%%%%%%%%%%%%%
\section*{References}
%%%%%%%%%%%%%%%%%%%%%%%%%%%%%%%%%%%%%%%%%%%%%%%%%%%%%%%%%%%%%%%%%%%%%%%%%
 
%%%%%%%%%%%%%%%%%%%%%%%%%%% References %%%%%%%%%%%%%%%%%%%%%%%%%%%%%%%%%%%% 

\end{document}